\documentclass[aps,prb,amsmath,amssymb,twocolumn,showpacs,superscriptaddress,final,floatfix]{revtex4-1}
\usepackage{graphicx}	
\usepackage{color}	
\usepackage[normalem]{ulem}
\usepackage{hyperref}\hypersetup{
  colorlinks,
  citecolor=blue,
  linkcolor=blue,
  urlcolor=blue}

\newcommand{\BFA}{BaFe$_2$As$_2$}

\newcommand{\eg}{{\em e.g.}}

\begin{document}

\title{Detailed band structure of twinned and detwinned \BFA~studied with angle-resolved photoemission spectroscopy}

\author{H. Pfau}
\email{hpfau@stanford.edu}
\affiliation{Stanford Institute for Materials and Energy Sciences, SLAC National Accelerator Laboratory, 2575 Sand Hill Road, Menlo Park, CA 94025, USA}
\affiliation{Department of Physics, Stanford University, Stanford, CA 94305, USA}

\author{C. R. Rotundu}
\affiliation{Stanford Institute for Materials and Energy Sciences, SLAC National Accelerator Laboratory, 2575 Sand Hill Road, Menlo Park, CA 94025, USA}
\author{J. C. Palmstrom}
\affiliation{Stanford Institute for Materials and Energy Sciences, SLAC National Accelerator Laboratory, 2575 Sand Hill Road, Menlo Park, CA 94025, USA}
\affiliation{Geballe Laboratory for Advanced Materials, Department of Applied Physics, Stanford University, Stanford, CA 94305, USA}
\author{S. D. Chen}
\affiliation{Geballe Laboratory for Advanced Materials, Department of Applied Physics, Stanford University, Stanford, CA 94305, USA}
\author{M. Hashimoto}
\author{D. Lu}
\affiliation{Stanford Synchrotron Radiation Lightsource, SLAC National Accelerator Laboratory, 2575 Sand Hill Road,
Menlo Park, California 94025, USA}

\author{A.F. Kemper}
\affiliation{Department of Physics, North Carolina State University, Raleigh, NC 27695, USA}

\author{I. R. Fisher}
\affiliation{Stanford Institute for Materials and Energy Sciences, SLAC National Accelerator Laboratory, 2575 Sand Hill Road, Menlo Park, CA 94025, USA}
\affiliation{Geballe Laboratory for Advanced Materials, Department of Applied Physics, Stanford University, Stanford, CA 94305, USA}

\author{Z.-X. Shen}
\affiliation{Stanford Institute for Materials and Energy Sciences, SLAC National Accelerator Laboratory, 2575 Sand Hill Road, Menlo Park, CA 94025, USA}
\affiliation{Department of Physics, Stanford University, Stanford, CA 94305, USA}
\affiliation{Geballe Laboratory for Advanced Materials, Department of Applied Physics, Stanford University, Stanford, CA 94305, USA}

\date{\today}

\begin{abstract}
We study the band structure of twinned and detwinned \BFA~using angle-resolved photoemission spectroscopy (ARPES). The combination of measurements in the ordered and normal state along four high-symmetry momentum directions $\Gamma$/Z--X/Y enables us to identify the complex reconstructed band structure in the ordered state in great detail. We clearly observe the nematic splitting of the $d_{xz}$ and $d_{yz}$ orbitals as well as folding due to magnetic order with a wave vector of $(\pi,\pi.\pi)$. We are able to assign all observed bands. In particular we suggest an assignment of the electron bands different from previous reports. The high quality spectra allow us to achieve a comprehensive understanding of the band structure of \BFA.

\end{abstract}

\maketitle

\section{Introduction}

In most high-temperature iron-based superconductors (SC), the superconducting transition temperature $T_c$ reaches its maximum when the nematic and magnetic phases are suppressed. \cite{paglione_2010} It was proposed that either nematic or magnetic quantum fluctuations may play a critical role for the high value of $T_c$. \cite{shibauchi_2014} Therefore, it is important to understand the impact of these two ordering phenomena on the electronic structure. In this paper, we will focus on the prototypical parent compound \BFA, but the results can be generalized to other 122 Fe-based SCs. 
 
The nematic phase transition breaks the rotational symmetry between $x$ and $y$ directions below $T_\mathrm{nem} = 137$\,K \cite{fernandes_2014}. It lowers the lattice symmetry from tetragonal to orthorhombic but is electronically driven \cite{chu_2012}. It leads to a different occupation of previously degenerate Fe $d_{xz}$ and $d_{yz}$ orbitals, which implies an energy shift of the corresponding bands in opposite directions. The nematic order develops twin domains. Mechanical stress and magnetic field have been used to obtain single-domain samples \cite{fisher_2011}. 
The spin-density wave (SDW) order with a wave vector of $(\pi,\pi,\pi)$ \cite{huang_2008} tracks the nematic one with $T_\mathrm{SDW} \lesssim T_\mathrm{nem}$ in the \BFA-family of compounds; they coincide by less than 0.3\,K in \BFA \cite{kim_2011_2}.  

Many angle-resolved photoemission experiments have been performed to study the evolution of the electronic structure across these two transitions. However, it is challenging to disentangle their influence, since $T_\mathrm{nem}$ and $T_\mathrm{SDW}$ are often very close.  The effects of the nematic order were recently studied in detail in FeSe \cite{tan_2013,shimojima_2014,nakayama_2014,watson_2015,zhang_2015,suzuki_2015,zhang_2016,fanfarillo_2016,watson_2016,fedorov_2016,watson_2017}, in which the magnetic order is absent. While the size of the nematic band splitting remains controversial \cite{zhang_2016,fedorov_2016}, a non-trivial momentum dependence could be extracted \cite{zhang_2016,suzuki_2015}. 
 
In the magnetically ordered 122 systems the situation is less clear.
The large number of bands in the ordered state in combination with a typically rather broad line-width poses a challenge to fully characterize the band structure in the ordered state. So far the following observations have been made on the 122 systems, in particular for \BFA: Many studies show a clear folding pattern due to SDW ordering  \cite{hsieh_2008,yang_2009,yi_2011_pnas,kim_2011,wang_2013,kondo_2010,liu_2009_prb,jensen_2011,yi_2009_prb,de_jong_2010,liu_2010_natphys,richard_2010,shimojima_2010,zabolotnyy_2009}. As a result one finds petal-like Fermi surfaces \cite{yi_2011_pnas,kim_2011,wang_2013,kondo_2010,yi_2009_prb,de_jong_2010,liu_2009_prb,liu_2010_natphys,richard_2010,shimojima_2010,zabolotnyy_2009,jensen_2011} and tiny Fermi surface pockets susceptible to Lifshitz transitions \cite{liu_2010_natphys,richard_2010}. Studies on de-twinned samples clearly observe a four-fold symmetry breaking attributed to the nematic order \cite{yi_2011_pnas,kim_2011}. Signs of band shifting and band splitting as expected for nematicity were observed \cite{wang_2013,liu_2009_prb,kondo_2010,jensen_2011}. For a specific momentum, a $T$-dependent nematic splitting could be extracted.\cite{yi_2011_pnas}. Simple calculations of the folded band structure are unsatisfactory to describe the experimental observations  \cite{kim_2011,yi_2009_prb}. This is unsurprising as we expect a non-trivial nematic splitting of the same energy scale as the SDW order parameter and the spin-orbit coupling. So far, a comprehensive band assignment in the ordered state of \BFA~is missing.
 
Here, we study the band structure of \BFA~in both ordered and paramagnetic states on twinned and detwinned samples using angle-resolved photoemission spectroscopy (ARPES). The high quality of the data enables us to identify a large number of details of the folded band structure not clearly resolved before. Since the magnetic ordering vector $(\pi,\pi,\pi)$ has a finite out-of-plane component, we combine measurements along four high-symmetry cuts $\Gamma$/Z--X/Y to determine the folding pattern. Starting from a clear characterization of the normal state, we are able to follow the band structure and perform a band assignment inside the ordered state.

\section{Methods}
 
High quality single crystals of \BFA~were grown using a self-flux method \cite{chu_2009,wang_2009,rotundu_2010}. ARPES measurements were performed at Stanford Synchrotron Radiation Lightsource beamline 5-2 with an energy resolution of better than 12\,meV and an angular resolution of 0.1$^\circ$. The base pressure stayed below $5\times 10^{-11}$\,Torr, and we compare results for $T = 20$\,K inside the ordered phase with 150\,K data above the transition temperature. The samples were cleaved {\it{in-situ}} below 30\,K. Linear horizontal (LH) and linear vertical (LV) polarizations are used to probe the different orbital characters \cite{zhang_2011_prb,brouet_2012,yi_2011_pnas}. The corresponding matrix elements are strongly momentum dependent \cite{yi_2011_pnas} and we indicate the expected dominant contributions of the $d_{xz}$, $d_{yz}$, and $d_{xy}$ orbitals for selected high-symmetry points in each spectrum.

\begin{figure}
\begin{center}
\includegraphics[width=\columnwidth]{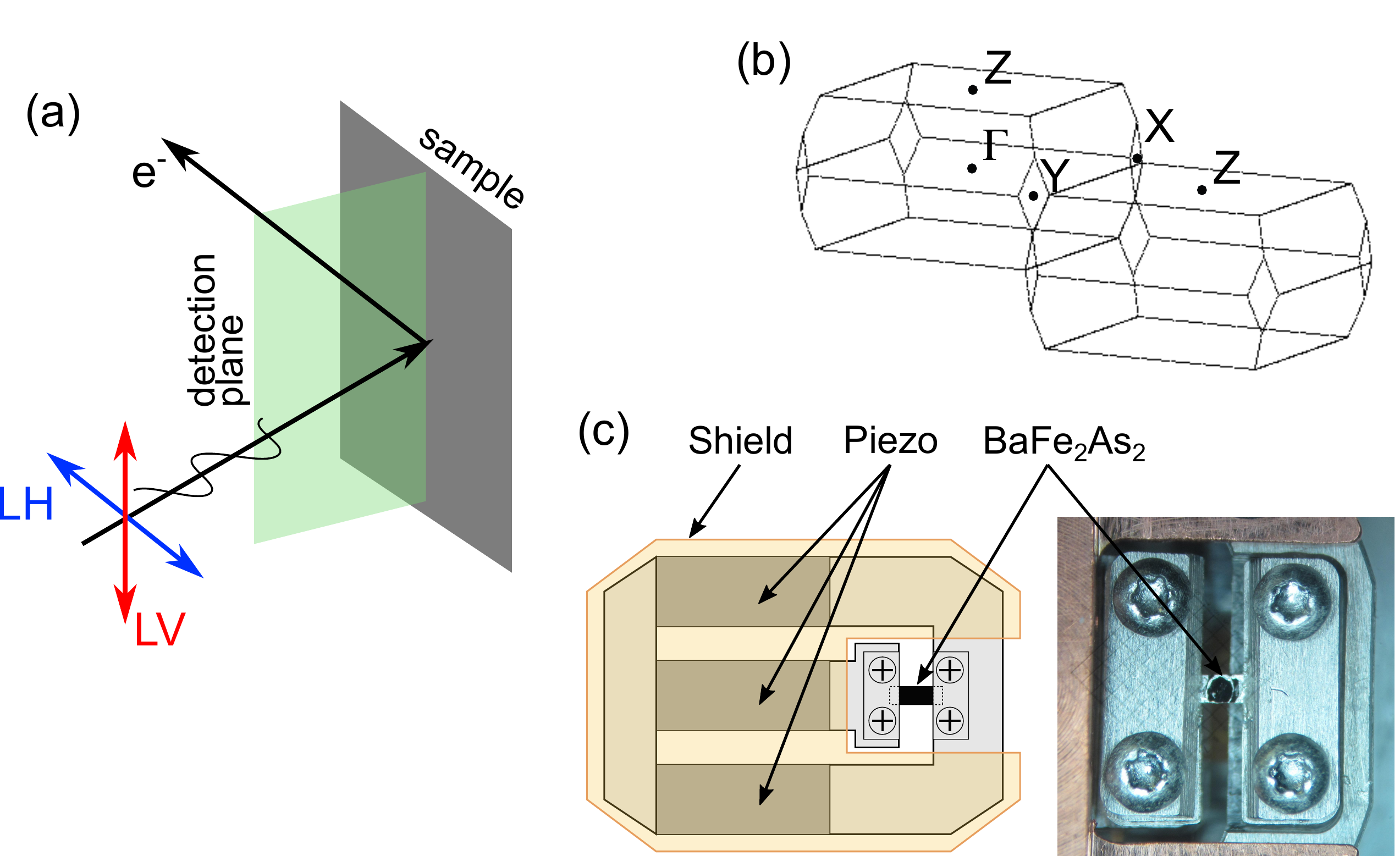}
\caption{
Measurement geometry. (a) Orientation of the sample, incoming photons and outgoing electrons. Linear vertical (LV) polarization is parallel to the analyzer slit and perpendicular to the sample normal. Linear horizontal (LH) polarization is perpendicular to LV polarization. (b) 2Fe-BZ of \BFA. The high-symmetry points are labeled according to the orthorhombic symmetry in the ordered state. (c) Sketch of the detwinning device and photograph of the mounted sample.
}
\label{Fig:geometry}
\end{center}
\end{figure}

Figures \ref{Fig:geometry}(a) and (b) sketch the measurement geometry and the notation in the orthorhombic Brillouin zone (BZ) of \BFA. All data were obtained using a photon energy of 47\,eV, which probes $k_z$ close to the $\Gamma$-plane in the first BZ and close to Z in the second BZ \cite{brouet_2009}. The high symmetry points X and Y are degenerate in the tetragonal phase. In the orthorhombic phase at low temperatures we studied both twinned and detwinned crystals. For twinned crystals, the typical beam spot size is on the order of 50\,$\mathrm{\mu m}$, much larger than the domain size \cite{chu_2010_prb}. Hence the ARPES signals integrate over both domains. To obtain the intrinsic single domain electronic structure, we detwin the single crystals using a device based on piezoelectric stacks as sketched in Fig.~\ref{Fig:geometry}(c), which is similar to Ref.~\onlinecite{hicks_2014}. The device is temperature compensated due to the arrangement of the piezoelectric stacks. The sample is detwinned by a compressive uniaxial pressure applied by the inner stack set to 200\,V. We verified that a metallic shield prevents the electric field from altering the ARPES measurements. The orthorhombic distortion of unstrained samples is $(a-b)/(a+b) = 0.49$\% far inside the ordered phase \cite{avci_2012}. The additionally applied strain to detwin \BFA~is $\Delta l/l < 0.02$\%, which we measured by a strain gauge. Since it is much smaller than the orthorhombic distortion, it has a negligible effect on the nematic band splitting far below $T_\mathrm{nem}$. 

We studied four different momentum directions $\Gamma$--X, $\Gamma$--Y, Z--X and Z--Y. To probe orthogonal directions in the detwinned samples, we rotate the sample with an in-plane rotation stage. The measurements in the $\Gamma$ plane were performed in normal emission corresponding to a light incident angle of 50$^\circ$ with respect to the sample normal. To reach Z--X/Y we rotated the sample towards smaller incident angles.


\section{Normal state band structure}
\label{sec:normal_state}

\begin{figure}
\begin{center}
\includegraphics[width=\columnwidth]{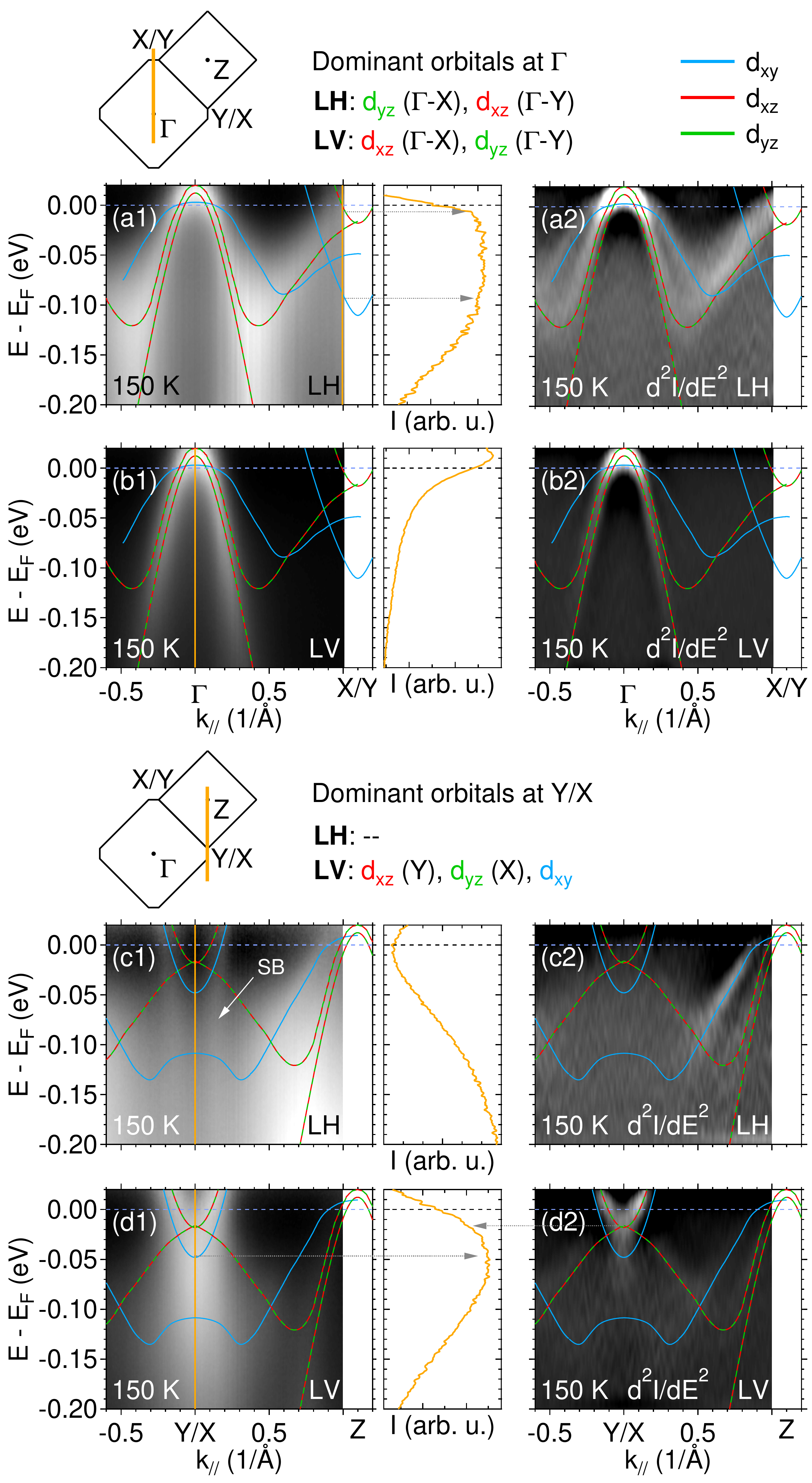}
\caption{
ARPES spectra of \BFA~at 150\,K. (a1,b1) Left: ARPES spectra divided by a Fermi-Dirac distribution taken along $\Gamma$--X/Y for LH and LV polarization, respectively. Right: EDC at the momentum marked with the yellow line. (a2,b2) Second derivative of the spectrum in (a1,b1). (c,d) same as (a,b) for momentum cuts along the Z--Y/X direction. Lines mark the band positions and are colored according to the orbital character.
}
\label{Fig:normal}
\end{center}
\end{figure}

\begin{figure}
\begin{center}
\includegraphics[width=\columnwidth]{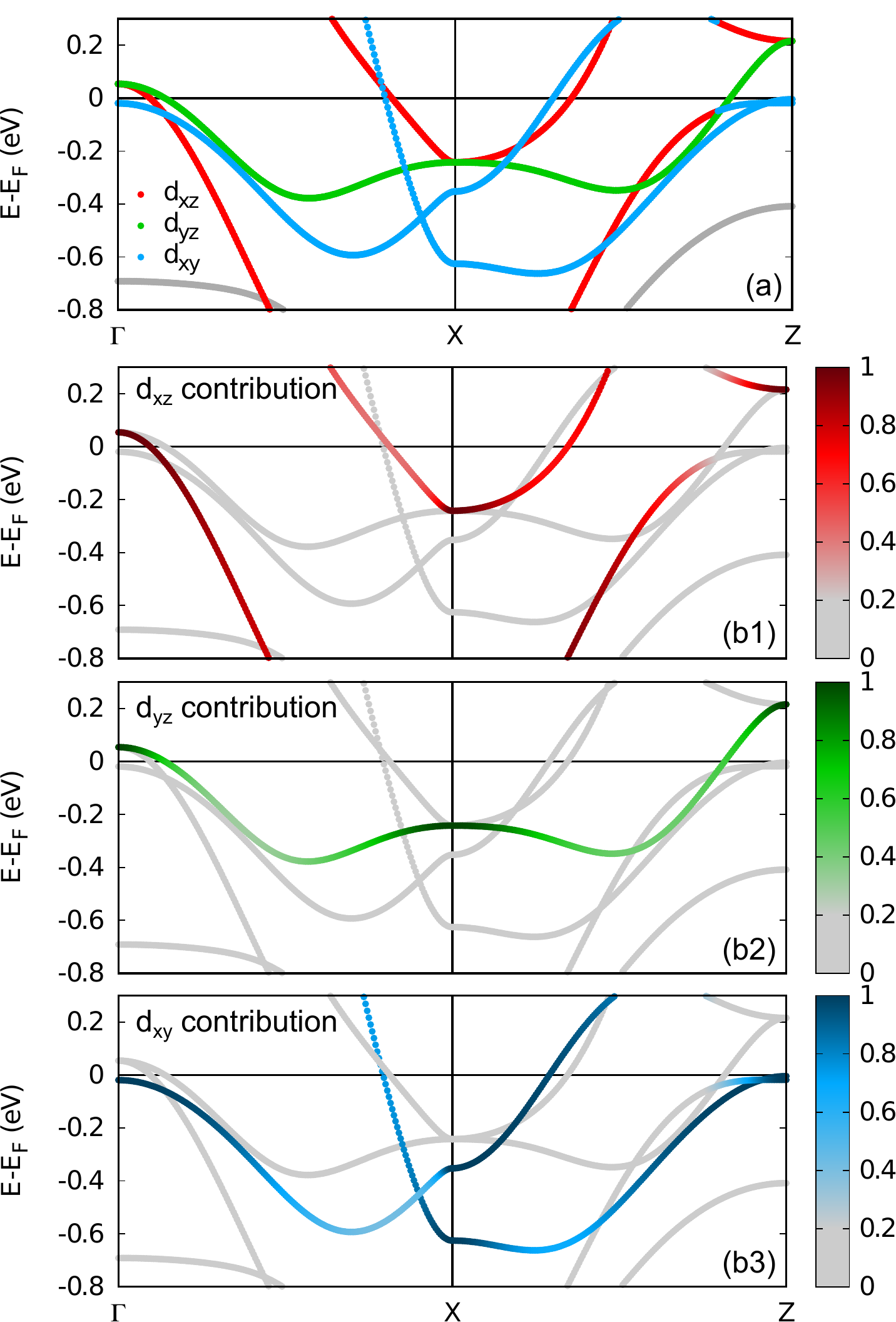}
\caption{
Band structure of \BFA~in the 2Fe BZ and in the normal state calculated from a tight binding model. We color-code bands according to their $d_{xz}$, $d_{yz}$ and $d_{xy}$ orbital character. We use (a) the majority character and (b) a color scale representing the precise orbital contribution. Along $\Gamma$--Y--Z, the dispersions are the same as shown here, but the orbital character changes from $d_{xz}$ to $d_{yz}$ and vice versa. Spin-orbit coupling, which is not included here, lifts the degeneracy of the $d_{xz}$ and $d_{yz}$ band at $\Gamma$ and Z but keeps it intact at X and Y. Note that the dispersion and orbital character of the $d_{xz}$ hole band along X--Z are affected by the hybridization with an electron band.
}
\label{Fig:LDA}
\end{center}
\end{figure}

\begin{figure*}
\begin{center}
\includegraphics[width=\textwidth]{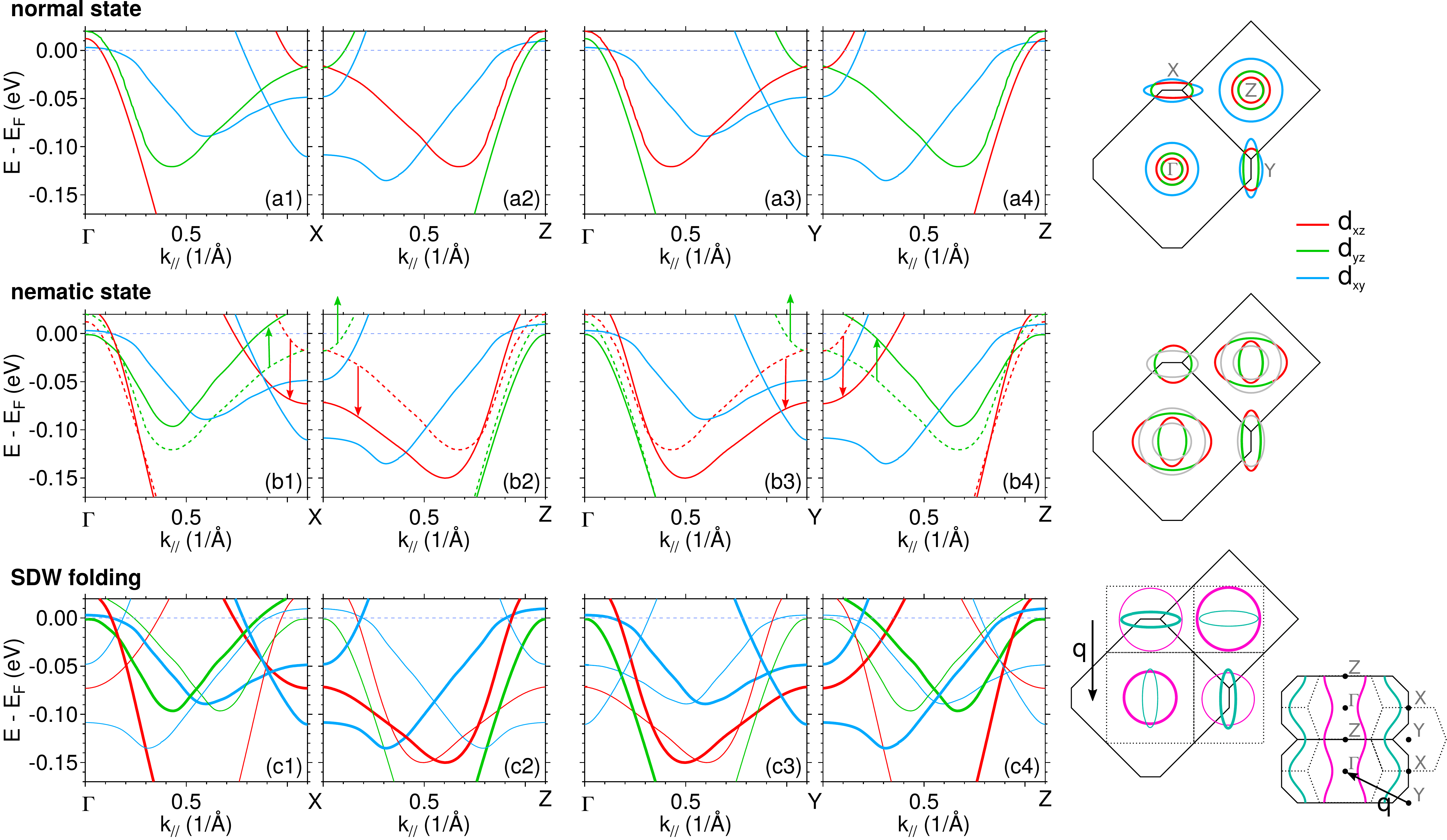}
\caption{
Illustration of the changes in the band structure due to nematicity and SDW order. (a) Left: Normal state band structure along four different momentum directions matching those studied in Figs.~\ref{Fig:normal},\ref{Fig:ordered_gamma}, and \ref{Fig:ordered_XY}. Binding energies are determined from ARPES spectra in Fig.~\ref{Fig:normal}. Right: Sketch of the Fermi surface. (b) Left: Orbital splitting due to nematicity. We include the normal state dispersion as dashed lines for comparison. We did not include effects due to spin-orbit interaction here. Right: Sketch of the Fermi surface distortion for the $d_{xz}$ and $d_{yz}$ bands. (c) Left: Bandstructure after SDW folding with a wave vector of $q = (\pi,\pi,\pi)$. We did not include the opening of SDW gaps. Right: Sketch of the folding scheme. For simplicity, we draw only one hole and one electron band. Thin lines represent folded bands.
}
\label{Fig:bands}
\end{center}
\end{figure*}

The band structure in the normal state of \BFA~(or doped versions of it) was studied numerous times \cite{mansart_2011,thirupathaiah_2010,zhang_2011_prb,yi_2009_prb2,jensen_2011,brouet_2012} and many of the features we will discuss below are consistent with previous studies. Here, we aim to obtain a complete data set along the high-symmetry cuts as a basis set for our discussion of the data in the ordered state, which are taken with the same photon energy. However, there are also certain aspects that extend or differ from previous reports. 

The ARPES results in the normal state at 150\,K are presented along the $\Gamma$--X/Y direction in Fig.~\ref{Fig:normal}(a,b) and the Z--Y/X direction in Fig.~\ref{Fig:normal}(c,d). We use LH and LV polarization to probe different orbitals and the second derivatives of the ARPES images to highlight band dispersions. We will use the band structure calculated from a tight binding model shown in Fig.~\ref{Fig:LDA} to assist with the band assignment of the observed dispersions and summarize the experimentally determined band structure in Fig.~\ref{Fig:bands}(a).

We identify the expected three hole pockets centered at $\Gamma$ [Fig.~\ref{Fig:normal}(a2,b2)] and at Z [Fig.~\ref{Fig:normal}(c2,d1,d2)] with $d_{xz/yz}$, $d_{yz/xz}$ and $d_{xy}$ orbital character. At X and Y we expect two electron pockets. They are more difficult to resolve and identify than the hole bands. We will discuss the assignment of the electron bands in detail in Sec.~\ref{sec:electron_bands}, where we consider both the high- and low-temperature data especially on the detwinned sample. The combination of all data sets will support the following discussion of the normal state. 

Along the Z--Y/X direction in Fig.~\ref{Fig:normal}(d) we find two electron bands: The energy distribution curve (EDC) at Y/X in Fig.~\ref{Fig:normal}(d1) shows a peak at $\sim 50$\,meV. The second derivative in Fig.~\ref{Fig:normal}(d2) highlights the existence of a second electron band with a band bottom at $\sim 20$\,meV at Y/X. The deeper electron pocket is predicted to be of $d_{xy}$ character, the shallower one is predicted to be of $d_{xz/yz}$ character (see Fig.~\ref{Fig:LDA}). Another deep electron band with a band bottom extending well beyond 200\,meV is visible in Fig.~\ref{Fig:normal}(c1). We assign this electron band to a surface band as will be further discussed in Sec.~\ref{sec:electron_bands}. Our assignment of the electron pockets along Z--Y/X differs from previous reports \cite{brouet_2012,jensen_2011,zhang_2011_prb}. The $d_{xy}$ and $d_{xz/yz}$ electron bands in Fig.~\ref{Fig:normal}(d) could not be separated in previous studies and were together assigned to one $d_{xz/yz}$ band. The deep electron band in Fig.~\ref{Fig:normal}(c1) was  assigned to the $d_{xy}$ band \cite{brouet_2012,jensen_2011} or $d_{x^2-y^2}$ \cite{zhang_2011_prb}.

We are not able to clearly observe the electron pockets along $\Gamma$--X/Y. For symmetry reasons, the $d_{xz/yz}$ band has the same binding energy at X/Y along $\Gamma$--X/Y as along Z--X/Y (see Fig.~\ref{Fig:LDA}). Therefore, we can assume a pocket of similar size, which we mark in Fig.~\ref{Fig:normal}(a,b). This assignment fits with the rise of intensity in the EDC of Fig.~\ref{Fig:normal}(a1) close to $E_\mathrm{F}$. The $d_{xy}$ electron band is predicted to have a lower binding energy along $\Gamma$--X/Y than along Z--Y/X (see Fig.~\ref{Fig:LDA}). This difference is due to the band folding from the 1Fe to the 2Fe BZ that contains a $k_z$ component in the body-centered tetragonal crystal structure of \BFA. We find a drop in the intensity at 100\,meV in the EDC close to X/Y [Fig.~\ref{Fig:normal}(a1)] and assign this signature to the $d_{xy}$ electron band. We will see later that this dispersion fits the corresponding folded electron band in the ordered state [Fig.~\ref{Fig:ordered_gamma}(d2)].

The middle and outermost hole bands are expected to extend towards the zone corner and be degenerate  with the electron bands at X/Y. This is consistent with the observations in Fig.~\ref{Fig:normal}(a2,c2,d2) and confirms the bottom of the $d_{xz/yz}$ electron band is at 20\,meV. The $d_{xy}$ hole band has a small matrix element close to X/Y [Fig.~\ref{Fig:normal}(a2)]. We therefore extend its dispersion such that it matches the $d_{xy}$ electron band at 50\,meV at X/Y as discussed above. This dispersion fits well the one observed in the ordered state (Fig.~\ref{Fig:ordered_gamma}(c2)), where orbital splitting moves the $d_{xz}$ band farther apart to uncover the $d_{xy}$ band. 

We can compare the experimentally determined band structure in the normal state shown in Fig.~\ref{Fig:bands}(a) with the calculated one from Fig.~\ref{Fig:LDA}. The bands are renormalized by a factor of approximately 3. However, a simple renormalization cannot reproduce the experimentally determined band structure. Instead, an orbital- and momentum-dependent renormalization and shift of the bands need to be considered, as is universally found in Fe-based SC \cite{yi_2017,lu_2008,charnukha_2015,lee_2012,brouet_2013,yin_2011,nishi_2011,richard_2011,yi_2009_prb2}. Density functional theory (DFT) calculations generally underestimate correlation effects, which in part gives rise to the observed renormalization. It has been suggested that the moderate correlation effects in Fe-
based SCs are driven by Hund’s rule coupling \cite{haule_2009,medici_2014}. On top of the overall renormalization, it was found that there is an 
orbital selectivity of the renormalization with increasing overall correlation strength \cite{medici_2014, yi_2017,yin_2011, yu_2013}. It is attributed to a decoupling of charge excitations between different orbitals which originates from Hund's rule coupling \cite{medici_2014, yin_2011, yu_2013}. The momentum-dependent shifts are less well understood. In principle, results from DFT calculations sensitively depend on the As height, which could lead to shifts. However, they cannot account for
the trend observed in Fe-based SCs. It was proposed that the strong particle-hole asymmetry in pnictides generates such momentum-dependent shifts \cite{ortenzi_2009}.


\section{Band structure in the ordered state}
\label{sec:ordered_state}

The band structure in the ordered state is modified by the nematic and SDW orders. The nematic order precedes the magnetic order by 250\,mK \cite{kim_2011_2}. The temperature difference becomes larger and more evident \eg~in Co-doped \BFA \cite{rotundu_2011}. In the nematic phase, the degeneracy between the $k_x$ and $k_y$ direction will be lifted due to rotational symmetry breaking. Therefore, we expect a difference in the binding energies of $d_{xz}$ and $d_{yz}$ bands along $\Gamma$--X and $\Gamma$--Y as well as along Z--X and Z--Y. In FeSe, the nematic band splitting is strongly momentum dependent and changes sign between the zone center and the zone corner \cite{zhang_2016,suzuki_2015}. We assume a similar momentum dependence here, which is confirmed by our recent studies on \BFA~and will be detailed in a separate paper \cite{pfau_to_be_published}. We sketch the expected band shifts in Fig.~\ref{Fig:bands}(b). We did not include gaps due to spin-orbit coupling in this sketch.

The SDW order has a commensurate ordering wave vector $(\pi,\pi,\pi)$ expressed in the basis of the tetragonal 2Fe BZ shown in Fig.~\ref{Fig:geometry}(b). It leads to a ferromagnetic ordering along the orthorhombic crystal axis $b$ and an antiferromagnetic ordering along the $c$ and $a$ axes ($b<a<c$) \cite{huang_2008}. The resulting folding pattern of the bands is sketched in Fig.~\ref{Fig:bands}(c). The folding in the magnetic state leads to the opening of SDW gaps, which we did not include in the sketch for better illustration. These gaps alter the dispersions and overlap with the signatures of the nematic band splitting and spin-orbit coupling. It is therefore difficult to unambiguously determine the nematic band splitting in the SDW phase.

The sketches in Fig.~\ref{Fig:bands}(b,c) will serve as a guide for the band assignment in the ordered state. The results from the ARPES measurements will, in turn, feed back into these sketches and allow for the estimation of the nematic band splitting.

\begin{figure*}
\begin{center}
\includegraphics[width=\textwidth]{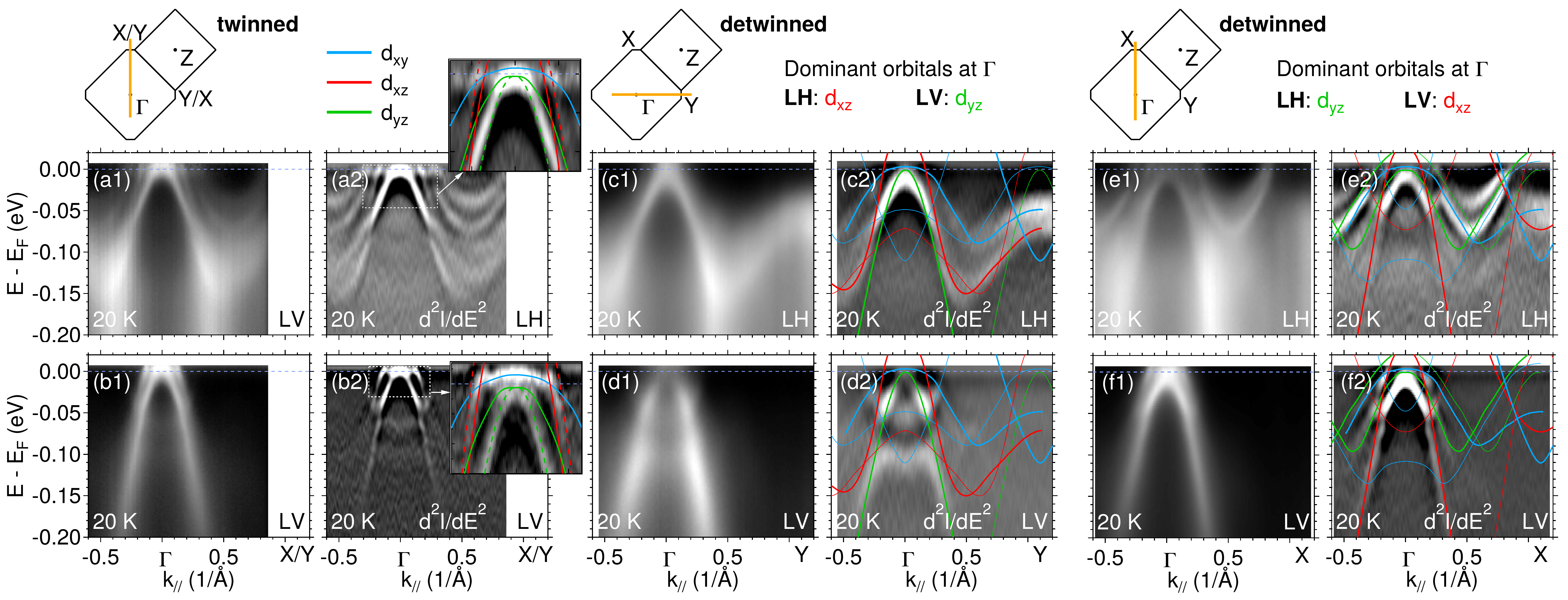}
\caption{
ARPES spectra close to the BZ center of \BFA~at 20\,K taken on (a,b) a twinned crystal and (c-f) a detwinned crystal. (a1,b1) ARPES spectra along $\Gamma$--X/Y divided by a Fermi-Dirac distribution for LH and LV polarization respectively. (a2,b2) Second derivative of (a1,b1). (c1,d1) ARPES spectra divided by Fermi-Dirac distribution along $\Gamma$--Y. (c2,d2) Second derivative of (c1,d1). (e,f) Same as (c,d) for momenta along $\Gamma$--X. Lines are taken from Fig.~\ref{Fig:bands}(c) and colored according to orbital character. Dashed and solid lines in (a2,b2) refer to the two nematic domains. Thin lines in (c2)-(f2) represent folded bands.
}
\label{Fig:ordered_gamma}
\end{center}
\end{figure*}

\begin{figure*}
\begin{center}
\includegraphics[width=\textwidth]{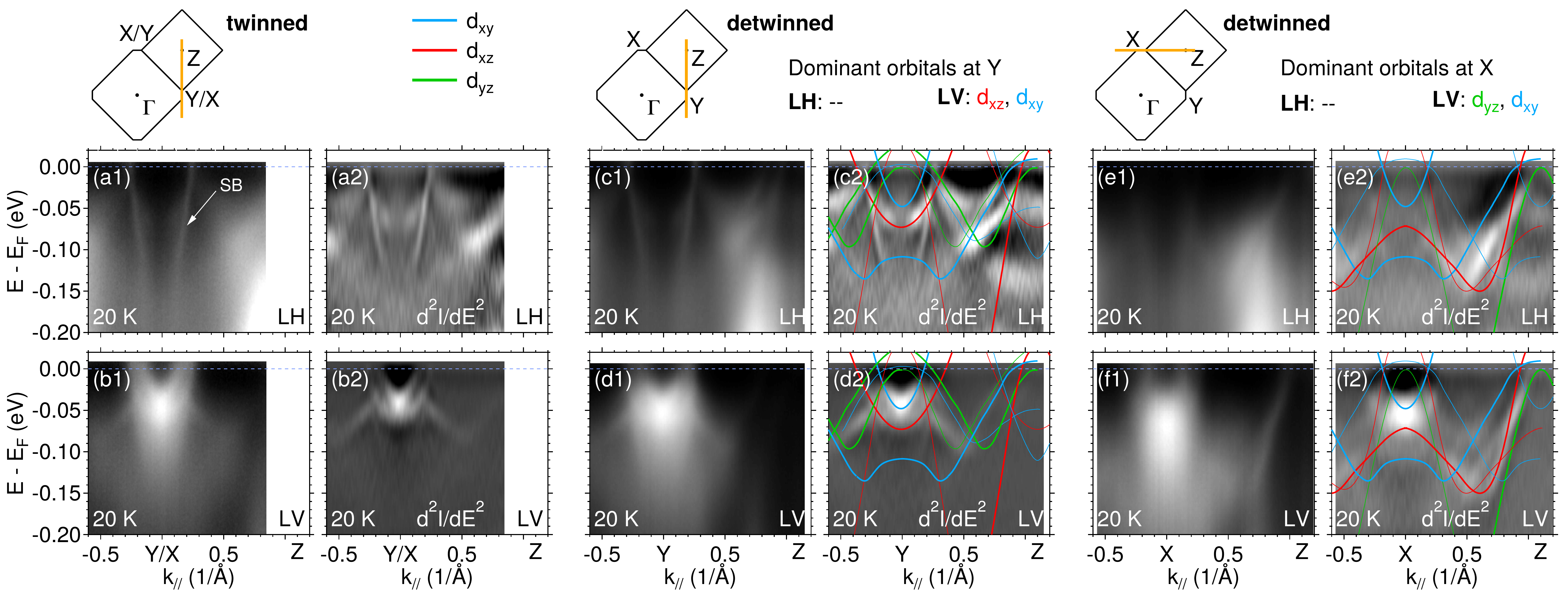}
\caption{
ARPES spectra close to the BZ corner of \BFA~at 20\,K taken on (a,b) a twinned crystal and (c-f) a detwinned crystal. (a1,b1) ARPES spectra along Z--X/Y divided by a Fermi-Dirac distribution for LH and LV polarization respectively. (a2,b2) Second derivative of (a1,b1). (c1,d1) ARPES spectra divided by Fermi-Dirac distribution along Z--Y. (c2,d2) Second derivate of (c1,d1). (e,f) Same as (c,d) for momenta along Z--X.  Lines are taken from Fig.~\ref{Fig:bands}(c) and colored according to orbital character.  Thin lines in (c2)-(f2) represent folded bands.
}
\label{Fig:ordered_XY}
\end{center}
\end{figure*}

Figure \ref{Fig:ordered_gamma} presents our ARPES results in the ordered state at 20\,K along the $\Gamma$--X and $\Gamma$--Y directions for twinned and detwinned crystals. Figure \ref{Fig:ordered_XY} summarizes our results along the Z--X and the Z--Y direction. We use again LV and LH polarization to probe different orbital contributions. We overlay the data on the detwinned sample with the folded band structure shown in Fig.~\ref{Fig:bands}(c). 

First, we will determine the nematic band splitting. From Fig.~\ref{Fig:ordered_XY}(d2) we can extract the shift of the $d_{xz}$ electron band at Y. In this spectrum, we find three concentric electron bands. The shallow one (band bottom: 50\,meV) has the same dispersion as the $d_{xy}$ band at 150\,K in the normal state [Fig.~\ref{Fig:normal}(d)]. The middle one (band bottom: 70\,meV) is therefore assigned to the $d_{xz}$ band that shifted down by 50\,meV due to the nematic order. The largest one (band bottom: $>200$\,meV) has the same dispersion as in the normal state and originates from a surface state. We will discuss this assignment in detail in Sec.~\ref{sec:electron_bands}. We do not find a signature of the $d_{yz}$ electron band in our measurements for the corresponding cut in Fig.~\ref{Fig:ordered_XY}(e2,f2). Assuming the same amount of shift for $d_{xz}$ and $d_{yz}$ bands in opposite direction, we expect it to be above the Fermi level [Fig.~\ref{Fig:bands}(b2,b3)]. The overall nematic splitting will then amount to approximately $100$\,meV at the zone corner for such a symmetric shift.

From Fig.~\ref{Fig:ordered_gamma}(c2) we extract the nematic band shift of the middle hole band with $d_{xz}$ character, which is now clearly separated from the $d_{xy}$ hole band. We can follow its dispersion for momenta $k_\parallel>0.4\,\mathrm{\AA{}^{-1}}$. The nematic band shift decreases away from Y. The influence of SDW gaps and spin-orbit coupling impedes the extraction of the exact band dispersion below $0.4\,\mathrm{\AA{}^{-1}}$. However, the very high quality of the spectrum in Fig.~\ref{Fig:ordered_gamma}(a2,b2) on the twinned crystal in conjunction with the detwinned data in Fig.~\ref{Fig:ordered_gamma}(c2)-(f2) helps us to identify the nematic shift of the hole bands around $\Gamma$. We find the $d_{yz}$ hole bands shifted just below the Fermi level. The $d_{xz}$ hole bands are shifted to higher binding energies at $\Gamma$ and we can identify their Fermi level crossing in the insets of Fig.~\ref{Fig:ordered_gamma}(a2,b2). Deviations from the observed dispersions can be explained by effects of spin-orbit coupling, which will hybridize the $d_{xz}$ and $d_{yz}$ bands.

The band positions of the $d_{xz,yz}$ hole and electron bands are used to obtain Fig.~\ref{Fig:bands}(b). We assume that the $d_{xz}$ and $d_{yz}$ bands move by the same amount in  opposite directions. The resulting folded band structure from Fig~\ref{Fig:bands}(c) fits very well the observed spectra in Fig.~\ref{Fig:ordered_gamma} and \ref{Fig:ordered_XY}. In Fig.~\ref{Fig:bands}(b,c), we omitted the opening of gaps due to spin-orbit coupling where $d_{xz}$ and $d_{yz}$ bands cross for better illustration. Gaps due to magnetic order are omitted for the same reason. It is particularly difficult to determine the size of the SDW gap as it is strongly orbital dependent \cite{plonka_2013,yi_2017}. This leads to certain discrepancies between the sketched band structure and the ARPES spectra particularly along $\Gamma$--X [Fig.~\ref{Fig:ordered_gamma}(e2)] and along Z--Y [Fig.~\ref{Fig:ordered_XY}(c2)], where the $d_{xy}$ and $d_{yz}$ hole bands follow each other very closely. However, we can exactly reproduce their number and roughly their binding energies. For all other bands we find very good agreement including binding energies.


\section{Electron Bands}
\label{sec:electron_bands}

\begin{figure}
\begin{center}
\includegraphics[width=\columnwidth]{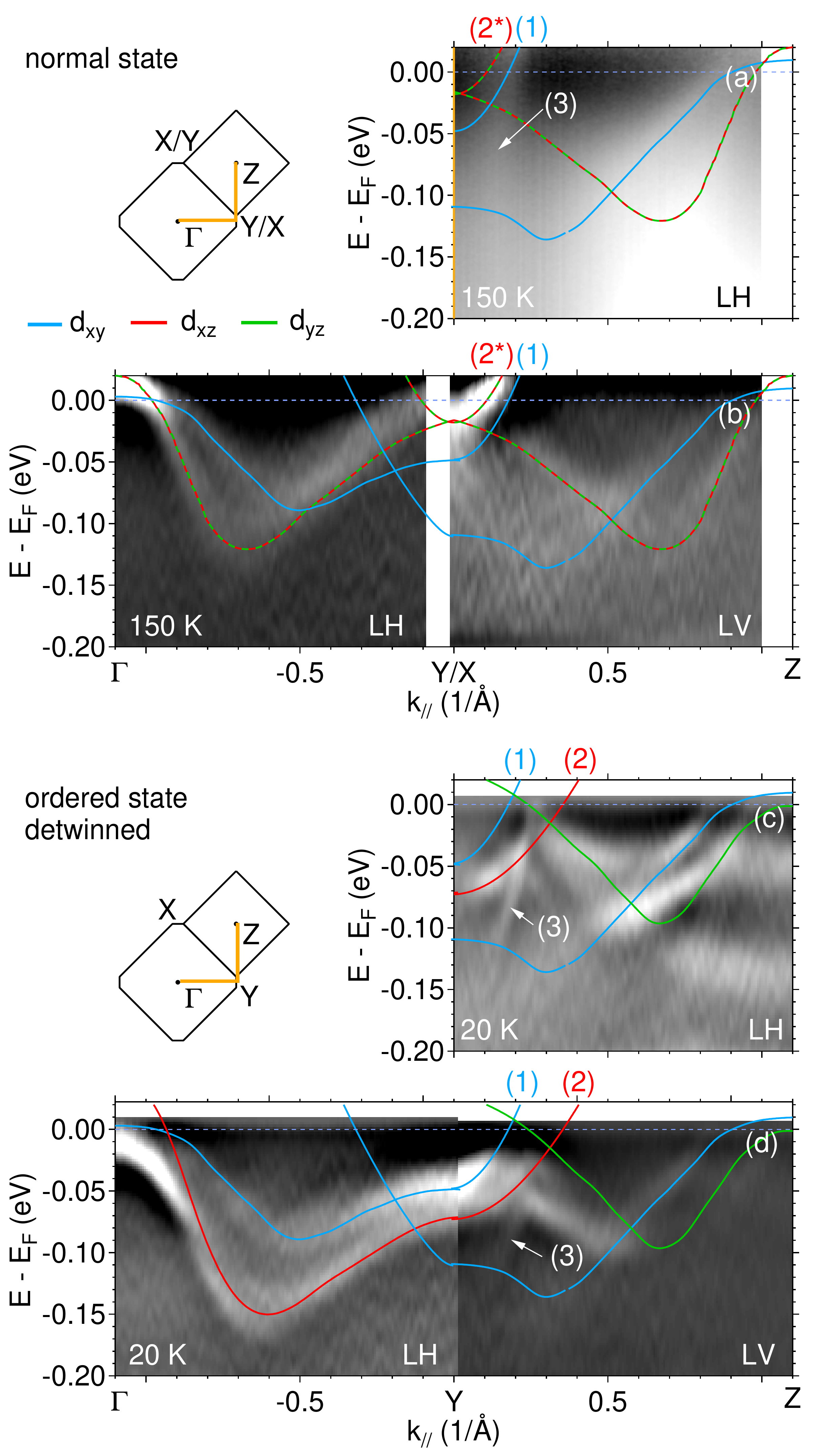}
\caption{
Assignment of electron bands. (a,b) Normal state spectra taken at 150\,K (a) along Y/X--Z with LH polarization and (b) along $\Gamma$--Y/X--Z with LH (left) and LV (right) polarization. (b) shows the second derivative spectra. (c,d) Second derivate of spectra taken in the ordered state on detwinned crystals (c) along Y--Z with LH polarization and (d) along $\Gamma$--Y--Z with LH (left) and LV (right) polarization. The spectra are taken from Figs.~\ref{Fig:normal},\ref{Fig:ordered_gamma}, and \ref{Fig:ordered_XY}. Lines highlight only specific bands important for the discussion of the electron bands. They are colored according to orbital character. 
}
\label{Fig:electron_bands}
\end{center}
\end{figure}

As we mentioned in Sec.~\ref{sec:normal_state}, our assignment of the electron bands differs from previous reports, specifically the bands marked (1, 2, 2*, 3) in Fig.~\ref{Fig:electron_bands}\cite{brouet_2012,jensen_2011,zhang_2011_prb}. All of them have been observed previously \cite{brouet_2012,yi_2009_prb,jensen_2011,zhang_2011_prb}. Band (3) was interpreted either as $d_{xy}$ \cite{brouet_2012,jensen_2011} or $d_{x^2-y^2}$ \cite{zhang_2011_prb}. Band (2) was assigned to the $d_{xz}$ orbital \cite{jensen_2011}. At low temperatures, the appearance of band (1) was interpreted as a surface effect \cite{jensen_2011}, while at high temperatures, bands (1) and (2*) could not be separated and were together interpreted as $d_{xz/yz}$. A detailed analysis of the electron bands in the ordered state on detwinned crystals was not reported previously. Its result indicates a band assignment as shown in Fig.~\ref{Fig:electron_bands} and a surface-related origin of band (3). The following arguments (i)-(vi) lead us to this assignment.

(i) In the spectra on detwinned crystals in Fig.~\ref{Fig:electron_bands}(c,d), we observe three electron bands (1,2,3) at Y where we expect only two. One of these bands is therefore likely a surface band. 

(ii) In principle, the binding energy of the $d_{xz}$ and $d_{xy}$ electron bands at Y is the same as the binding energy of the hole bands from the cut along $\Gamma$--Y. The lower hole band in Fig.~\ref{Fig:electron_bands}(d) has been identified as $d_{xz}$ \cite{yi_2011_pnas}. The upper hole band was observed before as well \cite{yi_2011_pnas,kim_2011} but never specifically assigned. Around $|k_\parallel| < 0.5\,\mathrm{\AA{}^{-1}}$ it has a dispersion similar to that of the $d_{xy}$ band in the normal state. At low temperature, we can now follow it all the way to Y and assign it to the $d_{xy}$ band. The comparison in Fig.~\ref{Fig:electron_bands}(d) shows excellent alignment of these two hole bands with electron bands (1) and (2). 

(iii) Since the $d_{xy}$ hole band shows no detectable change across the nematic transition, we assume that band (1) also has the same dispersion at low and high temperatures. Its band bottom agrees with the maximum in the EDC of Fig.~\ref{Fig:normal}(d1). Band (2*), highlighted by the second derivative in Fig.~\ref{Fig:electron_bands}(b), in turn must be the $d_{xz/yz}$ band. The remaining band (3) has to originate from a surface state. 

(iv) Another indication is the narrower line width of band (3) compared to those of bands (1) and (2), which is expected for a surface band. 

(v) Indeed, slab calculations predict a deep electron-like surface band at the zone corner \cite{heumen_2011}.

(vi) The orbital assignment agrees with the ARPES matrix elements for different light polarization considering effects due to folding from the 1Fe BZ to the 2Fe BZ \cite{brouet_2012,lin_2011} in addition to the momentum dependence of the matrix elements \cite{yi_2011_pnas}. Specifically, the $d_{xy}$ and $d_{xz}$ electron bands both appear in LV polarization along Z--Y (Fig.~\ref{Fig:electron_bands}(b,d)).

Previously, the binding energy of band (1) in the ordered state was found to be sensitive to temperature cycling and to be cleave dependent \cite{jensen_2011}. It was therefore assigned to a surface band.  The cleaving surface of \BFA~consists of half the Ba layer and is very sensitive to temperature cycling both by surface contamination and possible reconstructions. Since ARPES is a surface sensitive probe, such changes can influence the measured spectra, very often in unpredictable ways. Our assignment is mainly based on the symmetry argument that the hole and electron bands need to match at Y. 

The low-temperature data on detwinned crystals also help us to observe the $d_{xy}$ electron band along $\Gamma$--Y. It has been rather diffuse in the normal state, but can be clearly observed in Fig.~\ref{Fig:ordered_gamma}(d2) as a folded band at $\Gamma$. 

\section{Conclusion}

In summary, we were able to find a complete band assignment for the band structure of \BFA~in the normal and ordered states. The key to this assignment is the measurement of four momentum directions, $\Gamma$--X, $\Gamma$--Y, Z--X and Z--Y, for both twinned and detwinned crystals to obtain the band splitting in the nematic phase and the complete folding pattern of the magnetic phase.

We observed a momentum-dependent band splitting between the $d_{xz}$ and $d_{yz}$ bands due to nematic order. The largest splitting between the $d_{xz}$ and $d_{yz}$ bands of approximately 100\,meV was found around the BZ corner. Our data are consistent with a sign change of the nematic splitting but the influence of SDW folding and spin-orbit coupling with subsequent gap openings make it challenging to precisely determine the momentum dependence. Beginning from the normal state dispersion and applying i) a nematic band splitting and ii) a SDW folding along the folding vector $(\pi,\pi,\pi)$, we obtained a band structure that fits our observed spectra very well. Our discussion of the folded band structure does not include effects of spin-orbit coupling and of band gaps due to magnetic folding. These effects can explain the slight discrepancies between the band model and ARPES spectra in cases where binding energies of different bands are very close. We discuss in detail the electron bands near the BZ corner and demonstrate that the combination of measurements along different momentum directions is vital for their assignment.

\begin{acknowledgments}

We thank B. Moritz for stimulating discussions. H.P. acknowledges support from the Alexander von Humboldt Foundation. J.C.P. is supported by a Gabilan Stanford Graduate Fellowship and a NSF Graduate Research Fellowship (Grant No.~DGE-114747). This work was supported by the
Department of Energy, Office of Basic Energy Sciences, under
Contract No. DE-AC02-76SF00515. Use of the Stanford Synchrotron Radiation Lightsource, SLAC National Accelerator Laboratory, is supported by the U.S. Department of Energy, Office of Science, Office of Basic Energy Sciences under Contract No. DE-AC02-76SF00515.
\end{acknowledgments}


\bibliography{pfau_2018_prb}

\end{document}